# OPTICAL PROPERTIES (UV-Vis AND FTIR) OF GAMMA IRRADIATED POLYMETHYL METHACRYLATE (PMMA)

V. N. Rai[1*], C. Mukherjee[2*], Beena Jain[3],

[1]Synchrotron Utilization Section

[2]Optical Coating Laboratory

[3]Laser Boi-medical Application Section

Raja Ramanna Centre for Advanced Technology, Indore-452013 (India)

[*]Homi Bhabha National Institute, Training School Complex, Anushaktinagar, Mumbai-400094, India

(Corresponding Author: vnrai@rrcat.gov.in)

## ABSTRACT

The effect of gamma irradiation on the UV-Vis and FTIR spectroscopy of polymethyl methacrylate (PMMA) foils has been studied. A new absorption band is observed in the visible spectral range due to color centers induced in the gamma irradiated PMMA. This band shows maximum absorption (low transmission) for 10 kGy irradiation, which decreases and saturates after 50 kGy followed by a further increase at 500 kGy. The FTIR peaks show an increased absorption up to ~100 kGy irradiation, which reverses for higher doses. Broad band absorption is observed in FTIR spectra around 1600 and 3600 cm$^{-1}$ due to absorption of moisture in the irradiated samples. The reduction in the absorption intensity at 1718 cm$^{-1}$ in the irradiated PMMA (> 100 kGy) is found associated with the demerization of the carbonyl groups. An initial increase in the absorption of FTIR peaks with increase in the doses of irradiation is due to increased cross linking in the PMMA structure that is induced by the absorption of moisture. The demerization of carbonyl groups becomes dominant at higher doses of irradiation (~500 kGy).

## 1. Introduction

Polymers are an important class of materials having different applications in various fields of science and technology, particularly in optics, electronics, biotechnology, photonics and space research (Ismayil et al., 2010; Hossain et al., 2014; Ennis and Kaiser 2010; Fowzy et al., 2011; Megusar 1997; Ahmed et al., 2012; Lee et al., 1999; Shrinet et al., 1984). The study of changes in the properties of polymer material under the effect of ionizing radiation (electron and ion beam, x-ray and gamma radiation) has great importance due to its use in hard radiation environments such as particle accelerator, nuclear power plants, spacecrafts etc. (Kudo et al., 2009; Rai 1989; Subramanyam and Subramanyam 1987; Lin et al., 2003; Sung and Cho 2005; Lu et al., 2000; Tiwari et al., 2014; Singh et al., 2010). Normally, the high energy radiation loses its energy by inelastic interaction with the material, which produces cascade of secondary electrons within the solid microstructure of the polymers. These secondary electrons can initiate breaking of chemical bonds and generation of excited and ionized species or radicals through inelastic scattering by the surrounding molecules in the polymers. It has been found that various types of chemical changes take place in the polymers during the irradiation such as evolution of gases ($H_2$, CO, $CO_2$), formation of vacancy clusters, generation of color centers, main chain scission(C-C bond scission), creation of new double bonds, radical-radical combination (cross linking) etc (Ismayil et al., 2010; Hossain et al., 2014; Ennis and Kaiser 2010). Normally, the cross linking process produces new chemical bonds between two adjacent polymer molecules, which results in an increase in the molecular weight. On the other hand chain scission in the polymer molecules decreases its molecular weight. Both of these reactions produce a change in the physical properties of the polymers mainly in its optical and structural properties.

Various types of studies have been performed on different polymers after irradiating it with high energy ionizing radiations. Particularly, the effect of gamma irradiation on the PMMA (Fig. 1) has been studied extensively by many groups of researcher. The intrinsic viscosity measurements have been performed (Wall and Brown 1957) to determine the number of scissions of PMMA induced by gamma ray irradiation. PMMA have been [18] investigated to better understand the dependence of color change on the radiation dose (Knappe and Yamamoto 1970). The presence of volatile products in the gamma ray irradiated PMMA have also been reported earlier (Todd 1960; David et al., 1970). Ohnishi and Nitta (1959) have studied the generation of free radicals in the irradiated PMMA, where as Lin and Lee (1992) have observed the influence of methanol on the optical properties of irradiated PMMA. Lee et al. (1999) have studied LET (Linear Energy Transfer) effect on the cross linking and scission mechanism in PMMA during the irradiation. The changes in the optical properties are found to be related with the variation in its transmission due to the change either in the band gap or due to generation of color centers in the polymers (Fowzy et al., 2011; Ahmed et al., 2012; Kudo et al., 2009; Lu et al., 2000). Lu et al. (2000) reported the generation of color centers in PMMA after gamma irradiation, where the concentration of color centers was found decreasing with an increase in the annealing time at elevated temperature. The generation of color centers in the PMMA has been reported even using ultra short laser pulses (Samad et al., 2010). They found that the quivering motion of the free electrons in the electric field of the laser pulse displaces the atoms from their equilibrium positions creating free radicals and double bonds that coalesce into color centers. The effects of gamma irradiation on the acrylic acid/methyl methacrylate have been studied under the low dose of irradiation, which provided important information about its optical and electrical properties along with its thermal stability (Fowzy et al., 2011). The demerization of carbonyl groups, generation of new energy traps in the form of conjugated bonds, an increase in the water absorptivity and cross linking in Kapton-H have been reported after the heavy ion irradiation (Quamara et al., 2004; Garg and Quamara, 2006, 2007). No such study is found about the PMMA particularly in the case of gamma irradiation (low doses). This indicates that a careful and systematic investigation of optical properties of PMMA after gamma irradiation is needed in order to find the possibility of above processes taking place during the irradiation particularly the absorption of moisture and the condition for demerization of carbonyl groups in PMMA.

This paper presents the optical absorption and FTIR (Fourier Transform Infrared) spectroscopy of PMMA in the form of thin foils before and after gamma irradiation in order to study the mechanism of its degradation. Results have also been compared with the electron and ion beam irradiated PMMA films.

## 2.    Experimental

The Clarex PMMA (CLAREX from M/S Nitto Jushi Kogyo Co. Ltd. Tokyo, Japan) has been used for this study, because not much information is available about its behavior under the gamma irradiation. The flat foils of thickness 200 and 500 µm were used for this experiment after cutting it into small pieces of size 20 x 10 mm in order to irradiate them with different doses of gamma irradiation. Few samples of PMMA were irradiated at room temperature using $^{60}$Co (2490 Ci, Gamma chamber 900) source of gamma radiation having dose rate of 2 kGy/h. Optical absorption spectra of pristine and irradiated samples were recorded in transmission mode with a step size of 2 nm in the wavelength range of 200 - 1100 nm using a spectrometer (Cary 50 spectrometer, from Varian) at room temperature. FTIR spectra were recorded at a resolution of 4 cm$^{-1}$ in the range from 600 to 4000 cm$^{-1}$ using FTIR spectrometer (FTLA 2000MB 104, from ABB Bomem). FTIR spectra were recorded in reflection mode in order to avoid the saturation in the absorption peaks. It was decided after recording transmission mode FTIR spectra of PMMA foil, which provided saturated peaks. FTIR signal was averaged at 40 scan per sample in order to achieve a good signal to noise ratio.  It is important to mention here that we have used the terminology as an increase or decrease in the absorption corresponding to decrease or increase in the transmission/ reflection intensity in the optical/FTIR spectra in most of the places in the text.

The difference spectra of FTIR of PMMA recorded in reflection mode before and after gamma irradiation were obtained in order to find structural changes in the PMMA as a result of irradiation. The values of differential absorption or additional absorption (AA) spectra due to the irradiation are calculated based on the FTIR reflection spectra using the formula (Rai et al., 2010) as

$$\Delta k = \frac{1}{d} \ln \frac{R_1}{R_2} \qquad \text{-------------------------------------- (1)}$$

where d is the sample thickness and $R_1$ and $R_2$ are the values of reflected intensity at particular wavelength obtained from the spectra of PMMA recorded before and after gamma irradiation respectively. In fact $\Delta k$ is just a difference between linear absorption coefficient after and before irradiation, where bulk contribution cancels out and the resulting spectrum remains mainly due to the induced effects.

### 3.    Results and Discussion

### *3.1    UV-Visible Spectroscopy of Pristine PMMA*

The optical transmission spectra of pristine thin (200 µm) and thick (500 µm) foils of PMMA are shown in Fig. 2. Both types of samples show high transmission above ~400 nm wavelength, which is expected as PMMA finds wide applications as an optically transparent window material. Both the samples show strong absorption edge (decrease in transmission) below 400 nm. This strong absorption edge is produced due to the electronic excitation within the carbonyl chromophores (C=O) present in the PMMA structure (Fig. 1). The most intense absorption band observed in the spectra from 200-250 nm is found to be due to $\pi \rightarrow \pi^*$ transition in the C=O system. These features are common in the case of both types of samples. The thin PMMA foil shows another absorption band peaking at 300 and 340 nm. However, these peaks merge together and provide an enhanced saturated absorption peak from 286 to 362 nm in the case of thick PMMA foil, which may be due to an increased number density of PMMA molecules in the increased absorption path length. These peaks are attributed to the forbidden $n \rightarrow \pi^*$ transition that are less intense in comparison to $\pi \rightarrow \pi^*$ transition in the case of thin foil sample. However, in the case of thick foil, both the bands show broadening along with saturation in the absorption. The absorption band edge also shifts towards higher wavelength side in the case of thick foil, which seems to be due to the broadening of the peaks. These observations are in agreement with the earlier observations (Ennis and Kaiser 2010).

### *3.2    Effect of Gamma Irradiation on UV-Visible spectroscopy of PMMA*

Figure 3 shows the transmission spectra of thin foil of PMMA before and after irradiation by the gamma rays at the doses of 10, 50, 100 and 500 kGy. This shows that irradiated PMMA spectra have similar spectral behavior as pristine PMMA except an increased absorption and broadening in the absorption band from 200-250 nm as well as in the double peaks at 300 and 340 nm. Here very small variation was noted in the intensity of these absorption peaks after increasing the doses of irradiation from 10 to 500 kGy. A new absorption band is also observed from 384 to 523 nm in the visible spectral range of the spectrum for all the doses of irradiation. The amplitudes of new absorption bands are found nearly equal for 10 and 500 kGy irradiation, whereas it is comparatively less for 50 and 100 kGy. This indicates that the gamma irradiation induces some chemical and structural changes in the PMMA foils, which increases its optical absorption in the wavelength range from 200 to 375 nm as well as from 384 to 523 nm region. Similar changes in the spectral region of 200-375 nm have been reported by many groups in the cases of electron and the ion beam irradiation of the PMMA. Such changes in this spectral range has been attributed due to the irradiation induced formation of conjugated dienes (-C=C-)$_n$ group through unsaturation of the PMMA chain (Ismayil et al., 2010; Hossain et al., 2014; Ennis and Kaiser 2010). An increase in the density of dienes group in the sample can result in an increase in the absorption over the broad spectral range from 200-375 nm. However, the observation of new band from 384 to 523 nm is due to the presence of color centers in PMMA foil as a result of generation of some types of radicals, vacancies and/or due to trapping of electrons and holes during gamma irradiation. Similar absorption bands due to color center have been reported in the PMMA after gamma and high energy laser irradiation (Lu et al., 2000; Samad et al., 2010). Present results indicate that the 200 μm thin PMMA foil has very less variation in the absorption intensity of the color center peak for different doses of gamma irradiation ranging from 10 to 500 kGy indicating the occurrence of saturation like condition after 10 kGy irradiation. However, such absorption bands due to color centers in PMMA are not observed in most of the cases of electron and ion irradiated PMMA (Ismayil et al., 2010; Hossain et al., 2014; Ennis and Kaiser 2010).

In order to verify the occurrence of saturation in the absorption bands due to the color centers, the same experiment was repeated with ~500 μm thick PMMA foils and the results are shown in Fig. 4. It shows that the absorption due to color center band increases after 10 kGy irradiation followed by a small decrease in the absorption for the doses of 50 and 100 kGy, which further starts increasing with an increase in the doses of irradiation to 500 kGy. In this case (500 μm foil) the changes in the amplitudes of absorption peaks are found more in comparison to the 200 μm thin foil. The variation in the transmission (absorption) intensity of the color center peak (450 nm) with the doses of irradiation in the case of thin and thick PMMA foils are shown in Fig.-5 for a comparison. This clearly shows that the absorption first increases at 10 kGy and then slightly decreases and saturates for higher doses, which further starts increasing at ~500 kGy in the case of both the foils. However, thick foils show comparatively more absorption for all the doses of irradiation. At the same time an increased dose of irradiation (50 – 500 kGy) shows slight increase in the absorption (close to saturation) and the broadening in the lower wavelength bands (200-375 nm). Mostly these observations are similar to those observed in the case of thin PMMA foil (Fig. 3). These observations indicate that all the doses (50-500 kGy) are creating chemical and structural changes in PMMA foils that induces a saturation type of behavior in the absorption peaks (Characteristic absorption and color center bands). In both types of films, the absorption of color center band increases at 10 and 500 kGy irradiation that seems to be due to generation of radicals as a result of chain scission in the PMMA foils. However, a decrease and saturation in the absorption seems to be due to decrease in the number of radicals, which may be the result of cross linking in both the foils between ~50 and ~100 kGy irradiation. Similar increase and decrease in the absorption has been reported earlier in the absorption spectra of PMMA after ~70 MeV carbon ion irradiation (Singh et al., 2010). A regular variation (only increase in the absorption) has also been reported earlier for the color center peaks in the case of gamma irradiated PMMA (Lu et al., 2000). It seems that the cross linking between 50 to 500 kGy is taking place due to some kind of induced chemical changes in the structure of the pristine PMMA (Clarex) foils after gamma irradiation. An increase in the absorption of the lower wavelength bands (200-375 nm) is found correlated with an increase in the absorption of the color center band (382-600 nm) after the gamma irradiation, which seems to be due to same type of chemical changes. The increased amplitudes of the absorption bands (color center) in the case of thick foils is due to the availability of more number of molecules in thick PMMA foil for gamma induced chemical reactions.

Normally, all the ionizing radiations affects the band gap of the PMMA due to the creation of defects and/or carbon enriched clusters as a result of partial evolution of the hydrogen molecules at higher doses of irradiation (≥100 kGy) (Ismayil et al., 2010 and Fowzy et al., 2011). The presence of defects in the PMMA as a result of bond breaking might lead to the formation of lower energy state within the band gap. However, Fig. 3 and 4 do not show any significant change in its absorption edge centered at ~ 380 nm indicating a little effect on the band gap as well as on the activation energy in the present case for 10-500 kGy irradiations. Similar results have been reported earlier (Fowzy et al., 2011) in the case of gamma irradiated Acrylic acid/Methyl methacrylate copolymer films. They found that the irradiation process has no significant effect on both the band gap and the band tail width values even after the gamma irradiation up to ~100 kGy. It starts changing slowly after further increasing the doses of irradiation (≥ 100 k Gy). The thermal stability of such polymers also increases with an increase in the doses of gamma irradiation up to ~ 100 kGy and then it decreases for a further increase in the doses of irradiation beyond 100 kGy. It is reported that such observations are possible only due to an increased cross linking in the polymer foils up to ~100 kGy irradiation (Omastova et al., 1997; Fowzy et al., 2011; Lopergolo et al., 2000). The absence of any change in the band gap seems to be due to dominance of cross linking up to 100 kGy, where as further increase in the doses of gamma irradiation beyond 100 kGy is inducing a change in the band gap in the PMMA due to increased chain scission. However, they have not mentioned any reason for this cross linking. Normally, the cross linking and the chain scission occur simultaneously during the irradiation of the polymers whereas the relative magnitude of the cross linking (gelling) to the chain scission (degradation) depend mainly on the structure of the polymer (Banford et al., 1996; Lee et al., 1999; Andrej et al., 2005). In the present experiment also, the absorption intensity of the color center band first increases (~10 kGy) and then decreases and saturates for 50 and 100 kGy (Fig. 4) indicating about the probable dominance of cross linking in this range. It seems that at low irradiation (~10 kGy) the initial chain scission in the PMMA is creating various types of free radicals, which acts as color centers. The low number density of free radical may not be sufficient for the onset of cross linking resulting a decrease and saturation in the absorption for ~50 to ~100 kGy irradiations. It seems that there must be some other factor that induces the cross linking at such a low dose of irradiation. However, further increase in the doses of irradiation (~500 kGy) increases the absorption significantly in the thick PMMA foils due to further increase in the chain scission creating more number of free radicals, where as thin foils show nearly a saturation condition. The saturation is occurring due to an equilibrium between the chain scission and the cross linking in the range of 50-100 kGy for both the foils. Here, the availability of large number of PMMA molecules for the chain scission and the radicals for the cross linking are increasing the amplitude of the absorption in the case of thick PMMA in comparison to the thin foil. However, it is important to find out the corresponding changes in the structural properties (IR spectra) of the PMMA in order to find the reason for the onset of cross linking at low doses of irradiation as discussed above in the case of color center band.

### 3.3   *IR Spectroscopy of Pristine PMMA*

FTIR measurement of the PMMA foils before and after gamma irradiation were carried out in the reflection mode in order to find out the nature of modifications in its chemical bonding and structure. The FTIR spectra of the pristine and the gamma irradiated PMMA foils of thickness 200 and 500 µm are shown in Figs. 6 to 9. Table -1 shows the assignment of the absorption peaks observed in the spectra, which are compared with the absorption peaks assigned and reported in the literatures (Ismayil et al., 2010; Ennis and Kaiser 2010; Tiwari et al., 2014). The spectra of the gamma irradiated PMMA foils are found nearly similar to that of the pristine PMMA foils except changes in the amplitude and the locations of absorption peaks. The spectra show an intense C-H symmetric and asymmetric stretching peaks located between 2844 and 3022 cm$^{-1}$. The absorption peaks attributed to the C-H stretching vibrations occur from the methyl carbonyl and the chain methyl pendent groups as well as due to the main chain methylene (CH$_2$) groups. The C-H vibration modes of the pendent groups are influenced by the neighboring chains resulting in a broad peak, which is expected (Ennis and Kaiser 2010). An intense absorption band observed at 1718 cm$^{-1}$ is found due to carbonyl (C=O) stretching vibrations, which is found in agreement with the earlier reported results (Ismail et al., 2010; Ennis and Kaiser 2010; Tiwari et al., 2014). Many intense absorption bands are observed below 1500 cm$^{-1}$. Particularly, this region contains strong absorption features due to various (C-H) deformation modes of methyl subunit between ~1500 and 1350 cm$^{-1}$. The (C-C-O) stretching vibrations of the methyl carbonyl group are observed between 1260 and 1150 cm$^{-1}$, whereas (C-O-C) vibrations are found between 1190 to 1150 cm$^{-1}$. The absorption features below 1000 cm$^{-1}$ has been assigned due to C-H rocking modes and C-C skeletal mode (Ennis and Kaiser 2010).

### 3.4    *Effect of Gamma Irradiation on IR spectroscopy of PMMA*

It is observed that the absorption intensity of most of the peaks in the spectra increases after the gamma irradiation up to ~100 kGy followed by a decrease in its intensity at higher doses (~500 kGy) for both types of foils. Some of the peaks show a small shift in its location along with change in its amplitude, particularly the peak at 1718 cm$^{-1}$ due to carbonyl group shifts to 1724 cm$^{-1}$. The main peaks in the spectra below 1300 cm$^{-1}$ also show an increase in the absorption up to ~100 kGy irradiation. The absorption decreases and reaches close to or even less than the absorption of pristine PMMA for the irradiation of ~500 kGy (Figs. 6 to 9). The maximum increase in the absorption is noted for the peaks in the spectra around 1600 cm$^{-1}$ in the case of gamma irradiation for ~10 kGy (Fig. 6 & 8). However, such increase is noted in the case of all the spectra recorded after irradiation. In fact these peaks are due to the presence of H-O-H bending peaks observed around 1600 cm$^{-1}$. This observation is found associated with the presence of sharp free OH stretching peaks around 3600 cm$^{-1}$ (Fig. 7 and 9) along with a broad peak from 3000 to 3500 cm$^{-1}$ (Garg and Quamara, 2007). This broad band peak 3000-3500 cm$^{-1}$ is not clear in the main spectra, but it can be clearly seen in the difference spectra presented in the following text.    The absence of such peaks in the pristine PMMA indicates that gamma irradiation induces absorption of moisture in the PMMA samples. The presence of absorbed water in irradiated polymers has been discussed earlier (Zhiyong et al., 2002) particularly in Kapton -H after heavy ion irradiation (Garg and Quamara, 2007).  The amplitudes of peaks due to moisture are found maximum in the case of 10 kGy irradiation for both 200 (Fig. 6 & 7) and 500 µm (Fig. 8 & 9) thick PMMA foils. Both types of foils show many fold increase in the amplitude of the absorption peak in the bands between 1300-1570 cm$^{-1}$ after gamma irradiation, which seems to be due to interaction between H-O-H bending mode and C-H deformation modes of the methyl subunits (Ennis and Kaiser 2010). An increase in the absorption intensity for the peak at 1636 cm$^{-1}$ has also been noted, which is maximum for the gamma irradiation of ~10 kGy. This peak can be either due to C=C stretching vibration as a result of an increase in unsaturation or due to interaction of H-O-H vibration with some other nearby vibration, because this peak (1636 cm$^{-1}$) shows very small/ no absorption in the case of pristine PMMA.

Demerization of carbonyl groups (C=O) in polymers is an important effect of irradiation of high energy particles particularly ions (Davenas 1989; Garg and Quamara 2007). Normally, a decrease in the absorption intensity of this band in irradiated samples shows the decrease in the number of carbonyl groups as a result of bond breaking. It should show a proportional decrease with an increase in the doses of irradiation. In the present experiment absorption peak corresponding to C=O vibration in the pristine PMMA sample occurs at 1718 cm$^{-1}$, where absorption increases after ~ 10 kGy irradiation and then it starts decreasing accompanied by a small shift to 1724 cm$^{-1}$ for higher doses of irradiation. Figure 10 shows a comparison of variation in the transmission intensities of some of the important peaks at 986, 1144, 1636 and 1718 cm$^{-1}$ with doses of irradiation for both types of foils. The peaks at 986 and 1144 cm$^{-1}$ show an increase in the absorption (decrease in transmission) nearly up to 100 kGy gamma irradiation and then absorption starts decreasing for higher doses of irradiation for both types of foils (Fig. 10 (a) and 10 (b)). However, maximum increase in the absorption was noted for 1636 and 1718 cm$^{-1}$ peaks after 10 kGy irradiation followed by a decrease for higher doses in the case of both types of foils as shown in Fig. 10 (c) and 10 (d). Here maximum increase in the absorption seems to be due to interaction of H-O-H vibration as a result of enhancement in water absorption after gamma irradiation. It seems that the absorbed water gets itself attached with the carbonyl groups and the oxygen of the sub unit of methyl carbonyl group resulting in a cross linked structure (Garg and Quamara, 2007). This cross linking suppresses the demerization of carbonyl group as well as reduces the breaking of sub unit of methyl carbonyl groups along with C-C, C-O-C and C-C-O bonds nearly up to ~100 kGy. This is why most of the peaks show an increase in the absorption nearly up to ~ 100 kGy irradiation in comparison to the pristine foil. The effect of moisture absorption or cross linking (decrease in absorption) starts decreasing after 10 kGy. However, the decrease in peak absorption becomes high near ~ 500 kGy. This indicates that the demerization of carbonyl group and breaking of bonds in methyl carbonyl subunits are increasing and the absorption of moisture is decreasing with an increase in the doses of irradiation. This type of behavior in the optical and FTIR spectra of PMMA has been reported earlier in the case of 70 MeV carbon ion irradiation (Singh et al., 2010). But they have not mentioned any reason for this observation. However, most of the studies related with the electron beam irradiation of PMMA do not show any reversal in its nature with an increase in the doses of irradiation (Ismayil et al., 2010; Hossain et al., 2014; Ennis and Kaiser, 2010). It may be either due to comparatively higher doses of irradiation in the case of electron beam or due to difference in the structural properties of the PMMA sample.

In order to find reason behind such observation the difference spectra of pristine PMMA (200 μm) with that of 0 and 500 kGy irradiated samples are obtained using eq. -1 and is shown in Fig. 11 (a, b). The difference spectrum of 10 kGy irradiated sample (Fig. 11a) shows the presence of two dominant band due to H-O-H bending vibration around 1600 cm$^{-1}$ along with OH stretching vibrations around 3600 cm$^{-1}$ (in positive direction). Other main peaks are also showing increase in positive direction. Fig. 11b shows the difference spectrum of 500 kGy irradiated sample, which has comparatively small band due to moisture around 1600 and 3600 cm$^{-1}$ and the band due to CH stretching in positive direction along with new decreased peaks at 750, 1140, 1240 and 1724 cm$^{-1}$ (in negative direction). The presence of peaks in the negative direction indicates about the breaking of C-C, C-O-C, C-C-O bonds along with demerization of carbonyl groups respectively. These negative peaks are absent in the case of 10 kGy irradiation as the whole spectrum is in positive direction along with dominant band due to absorption of moisture. This confirms that the increase in moisture absorption after gamma irradiation increases the cross linking whereas bond breaking and demerization of the carbonyl groups become dominant at higher doses of irradiation (>100 kGy) in thin PMMA foil. However, the difference spectra of thick (500 μm) foil shows spectrum like Fig. 11a for 10 and 500 kGy irradiation without any demerization effect indicating that cross linking is dominant process even at 500 kGy in the case of thick PMMA. This shows that an increase in the absorption intensity of the peaks in comparison to the pristine samples up to 100 kGy gamma irradiation is due to the dominance of cross linking, whereas maximum increase in the absorption of the peak at 1636 cm$^{-1}$ and 1718 cm$^{-1}$ at 10 kGy irradiation is the result of strong band due to moisture. The absorption band at 1636 cm$^{-1}$ occurs due to the generation of C=C bond as a result of cross linking between the radicals after the initial chain scission. The absence of any significant change in the intensity of peak at ~2357 cm$^{-1}$ indicates about the negligible amount of $CO_2$ gas evolution in the present case. It seems that the lower dose (10 kGy) of irradiation induces absorption of sufficient amount of moisture in the PMMA foils generating the cross linking process. However, further increase in the doses of irradiation (500 kGy) induces a decrease in the amplitudes of absorption peaks due to dominant increase in the chain scission. Here maximum variation in the amplitudes of the absorption peaks is noted in the case of 200 μm PMMA foil in comparison to 500 μm thick foils (Fig. 10). Such changes seem to be related with the fraction of PMMA molecules affected by the doses of gamma irradiation resulting in either cross linking or the breaking of bonds in the polymer foil. However, establishment of equilibrium between chemical reactions (chain scission and cross linking) may be the reason for the observation of saturation like condition between 10 to nearly 500 kGy irradiation.

It has been reported that the structure of the polymer having more number of branching are more susceptible towards degradation/ chain scission (Chapiro, 1988). The PMMA structure is also having many branching, so it should have shown only the chain scission with an increase in the doses of irradiation. But in the present case of Clarex PMMA, results show the presence of both types of processes the cross linking as well as the chain scission for different doses of irradiation. It seems to be possible due to the nature of Clarex PMMA showing absorption of moisture after gamma irradiation. Here the presence of cross linking in the Clerax PMMA till ~100 kGy makes this material radiation resistant, which is beneficial for the nuclear industry and space applications.

### 3.5  *Polymer degradation*

Normally, the ionization and excitation phenomena occur in the polymers due to transfer of energy to the molecules of the medium during the irradiation process. Polymers endure bond cleavages after the absorption of energy resulting in the formation of non-saturated fragments called free radicals, which are responsible for most of the chemical transformations detected in the polymers (Chapiro, 1988). Various other processes take place in the polymers after irradiation such as cross-linking of polymer chains, chain-scission, oxidative degradation, changes in the unsaturation and the evolution of gaseous products etc. The magnitude of these processes relies upon the nature of the polymer, type of irradiation and the post irradiation conditions. In the present case of gamma irradiation of PMMA, no peaks due to $CO_2$ or CO is observed, which indicates about the absence of the chemical reactions as reported earlier in the case of gamma irradiation (Ennis and Kaiser, 2010). The presence of radicals (defects) in PMMA acts as color centers (new energy state) in the polymers (PMMA), which are found responsible for the emergence of a new absorption band in the optical absorption spectra of the PMMA. The emergence of color center peak and variation in its amplitude after gamma irradiation indicate about the generation of radicals and corresponding change in its number density due to chemical reaction in the polymer (PMMA) respectively. The decrease in the absorption of the color center peak between 10-100 kGy is due to onset of cross linking process. Normally, the presence of large number of primary radicals makes it react with each other resulting in either a recombination, cross-linking, end-linking or the creation of double bonds (C=C) (Chapiro, 1988). Mainly, the radical-radical reaction is responsible for the cross-linking. In the present case cross linking is taking place due to increased absorption of moisture in the PMMA after the gamma irradiation at low dose (10 kGy). In this case the absorbed water gets itself attached with the carbonyl group and the oxygen of methyl carbonyl sub unit resulting in a cross linked structure (Garg et al., 2001; Garg and Quamara, 2007). The cross-linkages are the covalent bonds, which modify the properties of the polymers where as the chain scission has a converse effect on the physical properties of the polymer. The chain scission is found dominant at higher doses of irradiation (~500 kGy) during the present experiment. Even in this case also the presence of moisture is noted. This indicates that the chain scission and cross-linking take place simultaneously during the irradiation, whereas the ratio of these two processes is dependent on the chemical structure of the PMMA as well as on the doses of irradiation.

## 4. Conclusions

The UV-Vis (optical) and FTIR spectroscopy of the PMMA foils (200 and 500 μm) have been investigated after gamma irradiation for the doses ranging from 10-500 kGy. The optical absorption spectra of both the foils show a new absorption band in the visible spectral range due to color centers as a result of generation of defects in the PMMA foils after the gamma irradiation. The absorption intensity of color center peak is found maximum at 10 kGy, which reverses (decreases) its nature for 50 and 100 kGy irradiations and further increases after increasing the doses of gamma irradiation to 500 kGy. FTIR spectra show an increase in the absorption intensity of nearly all the peaks up to ~100 kGy in comparison to the pristine PMMA, which decreases after further increase in the doses of irradiation. Difference spectrum of gamma irradiated (10 kGy) and pristine sample shows the presence of new absorption bands due to OH and H-O-H vibration. This shows that the absorption of moisture is taking place in PMMA after ~10 kGy irradiation that induces cross linking process in PMMA resulting in a decrease in the absorption of FTIR peaks. The decrease in the absorption of color center band between 10 to 100 kGy is also due to this cross linking. The difference spectrum of 500 kGy irradiated and pristine sample shows a comparatively small increase in the absorption due to moisture along with drastic decrease in the absorption due to carbonyl and methyl carbonyl group subunit. This is due to demerization of carbonyl group at higher doses of irradiation. This indicates that absorption of moisture is taking place at low doses of irradiatation, which induces cross linking in the PMMA, where as chain scission becomes dominant at higher doses of irradiation. To the best of our knowledge, we have observed first time the absorption of moisture by PMMA after gamma irradiation that induces cross linking between ~10 to ~100 kGy irradiation. Finally, these results show that this PMMA sample behaves as a radiation resistant material till ~100 kGy, which may be beneficial in the nuclear industry as well as in the space applications.

**Acknowledgment**

The authors are grateful to Vishal Dhamgaye for providing the PMMA foils and S. Kher for irradiating it for this experiment. Authors are thankful to P. A. Naik for his interest in this work.

Table -1

FTIR peak assignment for PMMA

| Peak assignment | Ref-3 | Ref-1 | Ref-15 | Present |
|---|---|---|---|---|
| C-C stretch | 751 | 750 | 751 | 750 |
| (C-C skeletal mode) | | | | |
| $CH_2$ rock | 840 | 844 | 840 | 841 |
| O-$CH_3$ rock | 910-999 | 989 | 989 | 986 |
| C-O-C asymmetric stretch | 1059-1401 | 1147 | 1170 | 1144 |
| C-C-O symmetric stretch | | 1241 | 1236 | 1240 |
| C-H deformation | | 1387 | 1385 | 1387 |
| (C-H bending vibration) | | | | |
| C-H deformation | 1426-1496 | 1442 | 1440 | 1437 |
| (C-$CH_3$ bending vibration) | | | | |
| $CH_2$ bending vibration | | 1484 | | 1489 |
| C=C stretching vibration | | 1636 | 1633 | 1635 |
| C=O stretching vibration | 1711 | 1728 | 1730 | 1718/1724 |
| C-H stretching | 2844-3022 | 2448 | 2848 | 2853 |
| (C-$H_2$ stretching vibration) | | | | |
| C-H symmetric stretching | | 2951 | | 2950 |
| C-H asymmetric stretching | | 2995 | 2998 | 2995 |

**Figure Caption**

1. Chemical structure of polymethyl methacrylate (PMMA).

2. Optical transmission spectra of pristine PMMA foils of two different thicknesses.

3. Optical transmission spectra of thin PMMA foil (200 μm) after gamma irradiation, for different doses.

4. Optical transmission spectra of thick PMMA foil (500 μm) after gamma irradiation, for different doses.

5. Variation in the optical transmission intensity at 450 nm with respect to doses of gamma irradiation in PMMA foils of different thicknesses.

6. FTIR spectra of thin PMMA foil (200 μm) recorded in reflection mode (500 - 2000 cm $^{-1}$) before and after gamma irradiation for different doses.

7. FTIR spectra of thin PMMA foil (200 μm) recorded in reflection mode (2500 - 4000 cm$^{-1}$) before and after gamma irradiation for different doses.

8. FTIR spectra of thick PMMA foil (500 μm) recorded in reflection mode (500 - 2000 cm $^{-1}$) before and after gamma irradiation for different doses.

9. FTIR spectra of thick PMMA foil (500 μm) recorded in reflection mode (2500 - 4000 cm $^{-1}$) before and after gamma irradiation for different doses.

10. Variation in the FTIR intensity (Reflectance) with respect to doses of gamma irradiation in PMMA foils of 200 and 500 μm at different infrared frequencies. (a) 986 cm$^{-1}$ (b) 1144 cm$^{-1}$ (c) 1636 cm$^{-1}$ (d) 1718 cm$^{-1}$

11. Difference spectra of PMMA foil (200 μm) before and after gamma irradiation. (a) 10 kGy (b) 500 kGy

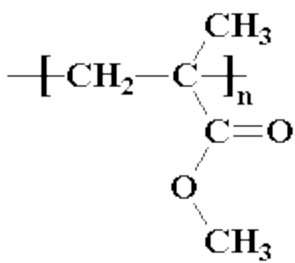

**Fig-1**

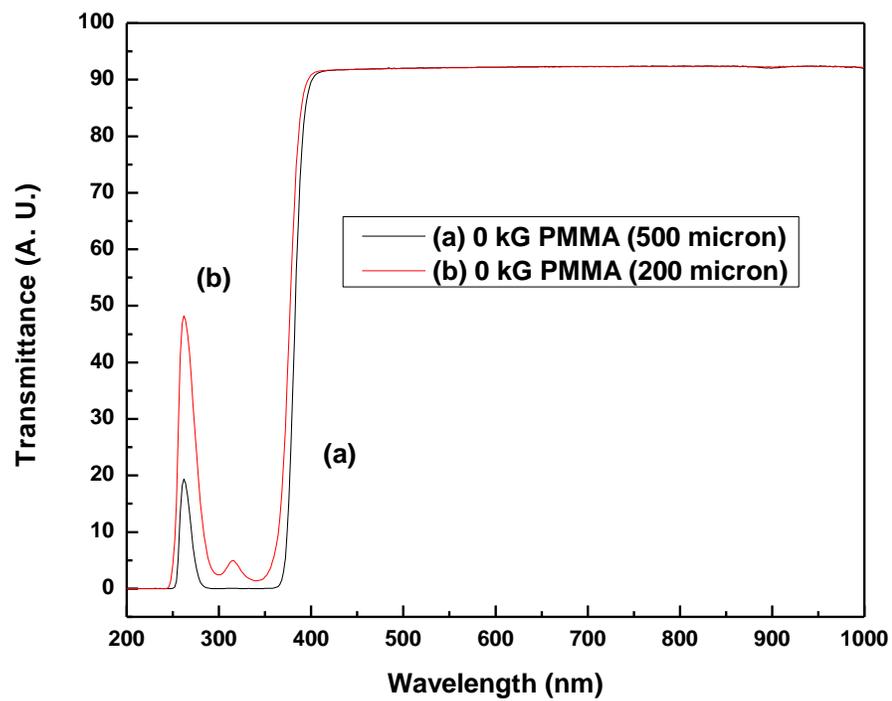

**Fig.-2**

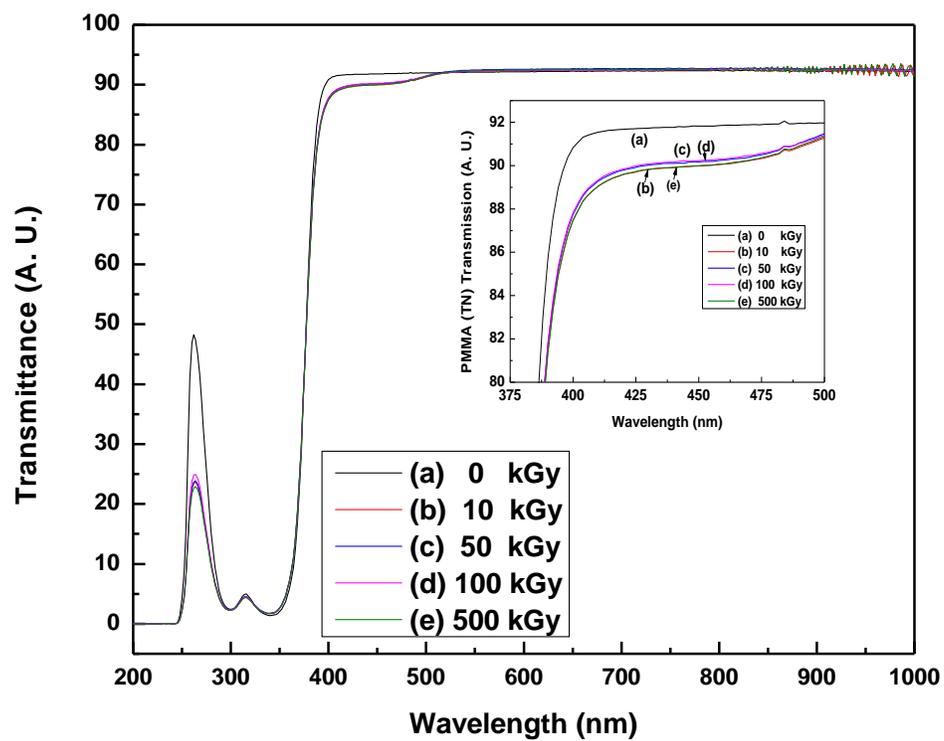

**Fig.-3**

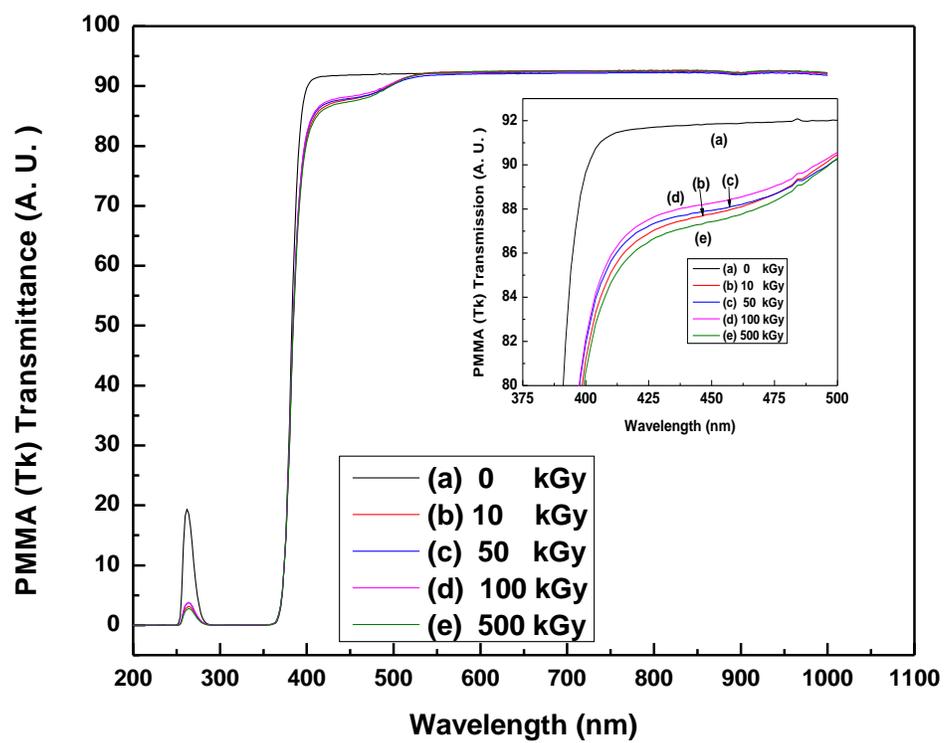

**Fig.-4**

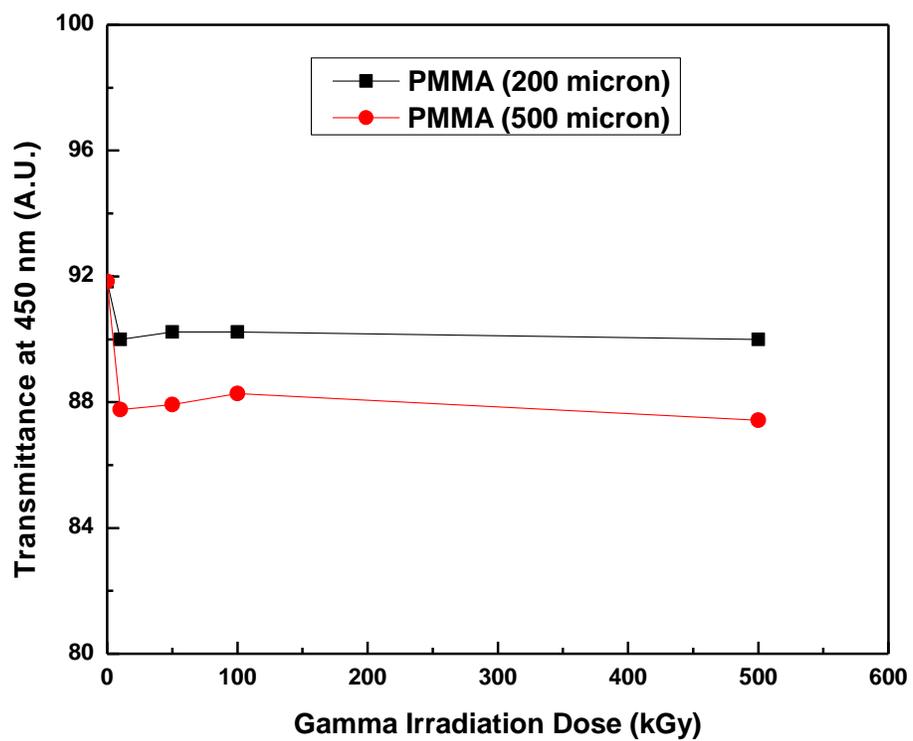

**Fig. 5**

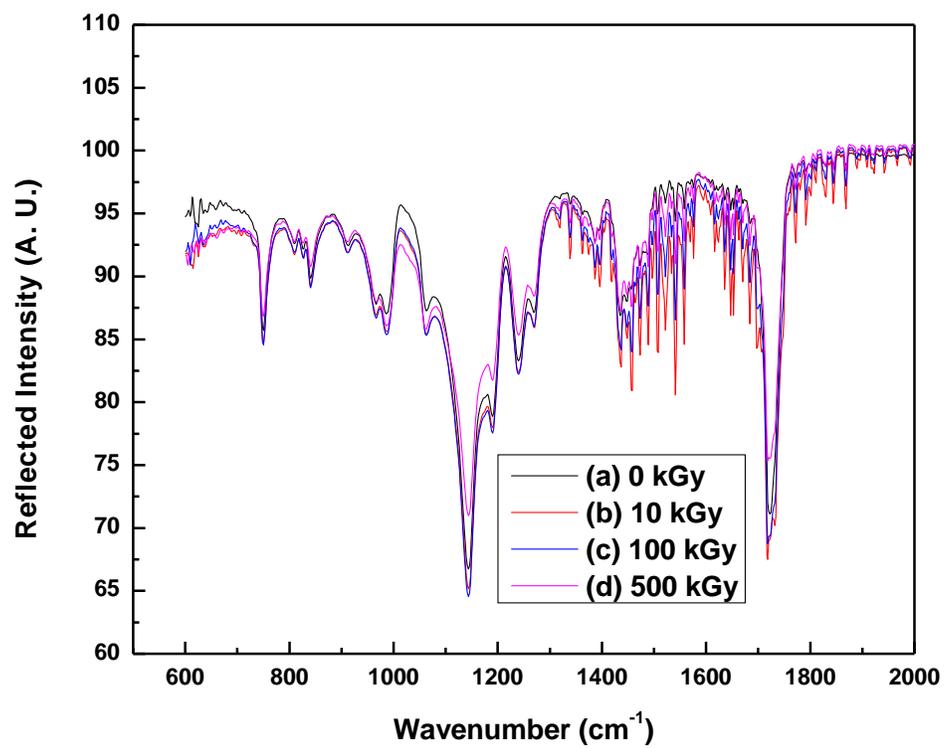

**Fig.-6**

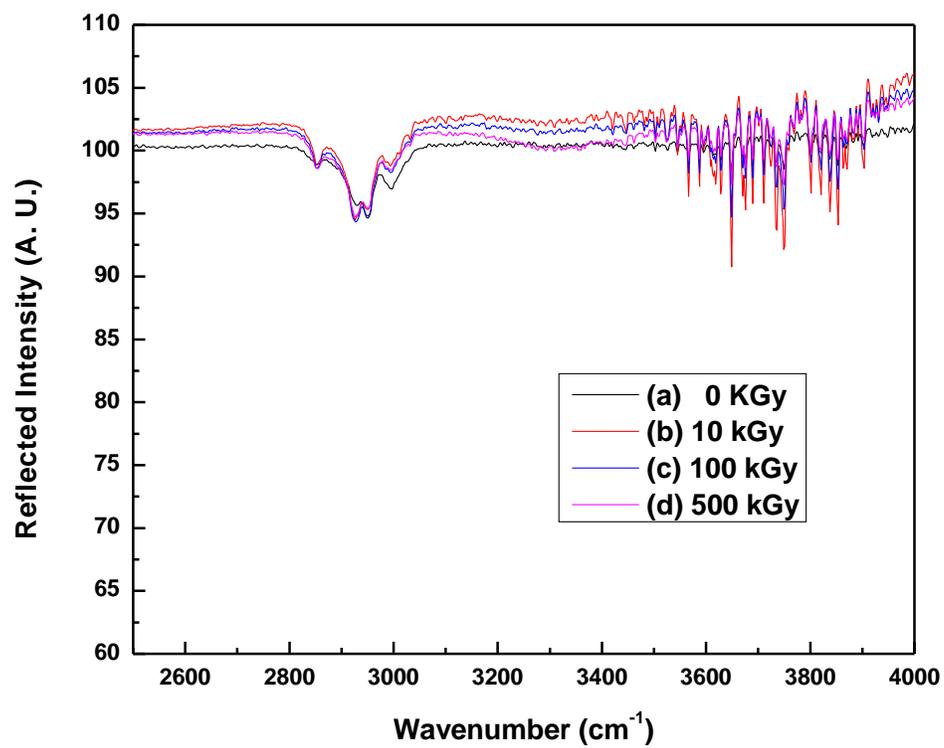

**Fig.-7**

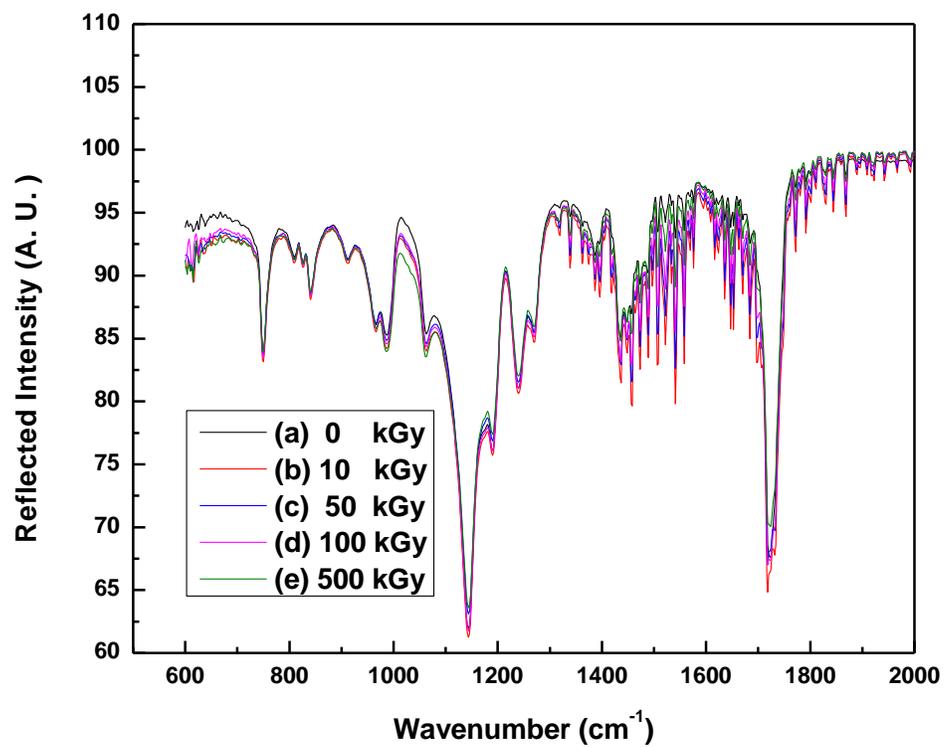

**Fig.-8**

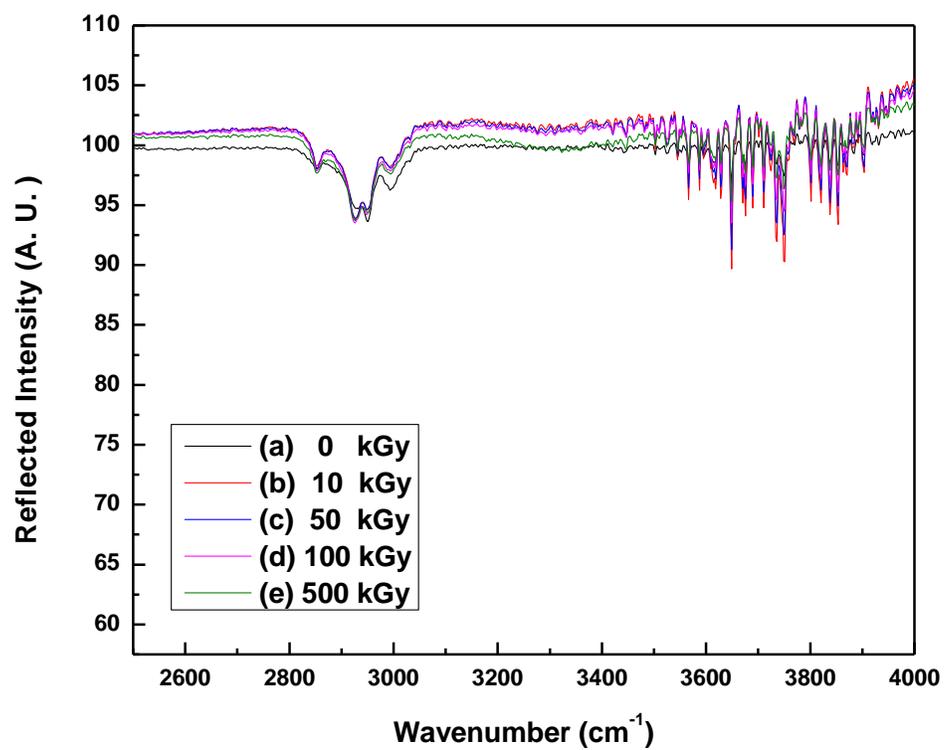

**Fig.- 9**

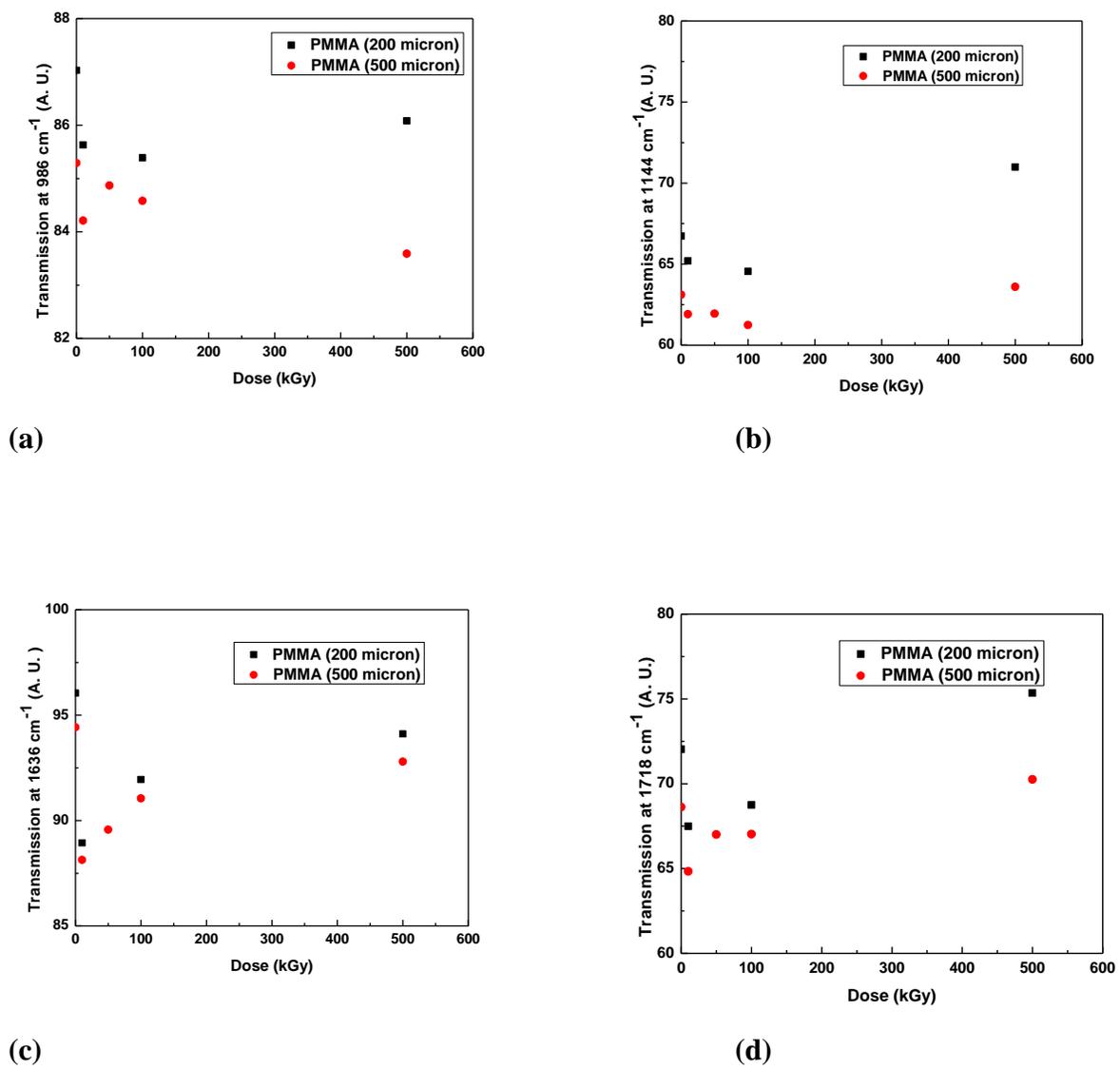

**(a)**

**(b)**

**(c)**

**(d)**

**Fig. 10**

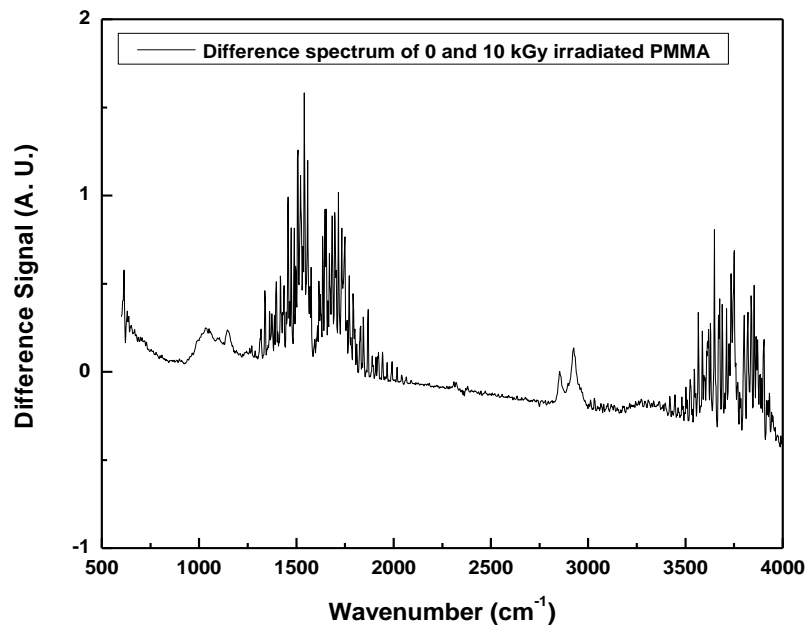

**(a)**

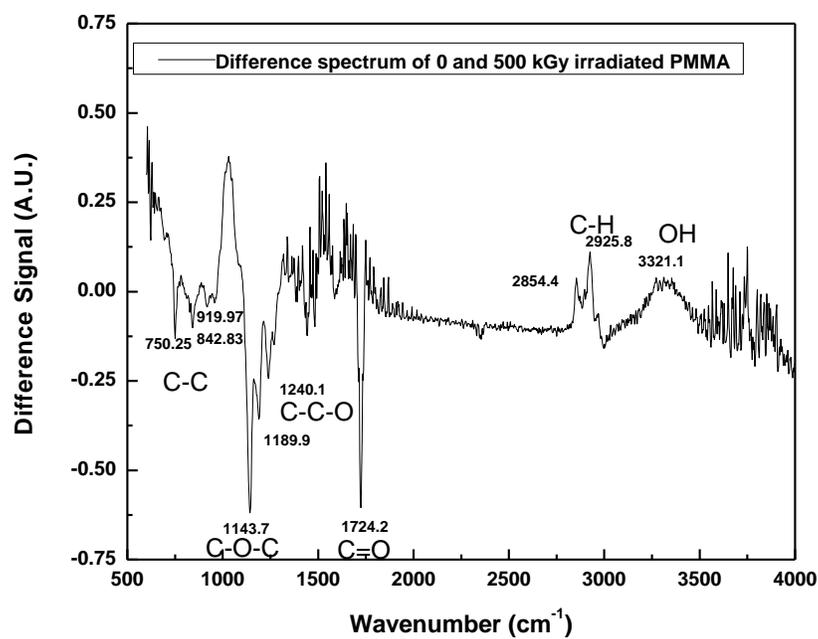

**(b)**

**Fig.- 11**